# Phonons in lattices with rod-like particles

A. Sparavigna

**Dipartimento di Fisica, Politecnico di Torino, C.so Duca degli Abruzzi 24, 10129 Torino, Italy**

The paper studies the modes of vibrations of a lattice with rod-like particles, in a continuum model where the sites of the lattice are the connections among strings and rigid rods. In these structures then, translational and rotational degrees of freedom are strongly coupled. We will discuss in particular two-dimensional lattices with auxetic-like behaviour. Auxetics are materials with a negative Poisson elastic parameter, meaning that they have a lateral extension, instead to shrink, when they are stretched. We assume as "auxetic-like" two-dimensional structures, structures which do not collapse, when stretched in one of the in-plane directions. The presence of rigid rod-like particles in the lattice prevents the shrinking of the membrane. Complete bandgaps between acoustic and optical modes are observed in analogy with the behaviour of crystalline materials.

**Introduction**
In this paper we discuss the vibrations of two-dimensional structures composed of flexible and rigid parts. The membrane contains rod-like particles connected with strings. In these structures, translational and rotational degrees of freedom are strongly coupled. As the models demonstrate, these structures are able to display complete band-gaps. Elastic materials can have phononic bandgaps - intervals of frequencies where no propagating phonons exist - in analogy with the photonic materials, known as photonic crystals, displaying bandgaps for light waves [1]. Phononic band-gaps can be found in crystalline materials [2,3] or in macroscopic periodic elastic media, the "phononic crystals" [4].

Another family of interesting materials are those with the auxetic elastic behaviour, characterised by a negative Poisson elastic parameter [5]: this relevant property means that these materials have a lateral extension, instead to shrink, when they are stretched. Natural auxetic materials and structures occur in biological systems, in skin and bone tissues and in crystalline membranes [6-8]. Man made auxetic materials have been produced from the nanoscale till the micro- and macroscales and cover the major class of materials, such as metals, ceramics, polymers and composites [9-14].

We propose as "auxetic-like" two-dimensional structures, those structures which do not collapse, when stretched in one of the in-plane directions. The mesh is composed of strings and rigid parts: it is the presence of rigid extended particles that prevents the shrinking of the membrane. The approach we follow in the calculations is that recently proposed by Martinsson and Movchan [15]; these authors studied the phonon dispersions for membrane-like lattices with several geometries and non-uniform mass distribution along the strings, finding phononic band-gaps.

**Flexible and rigid parts in the mesh.**
To understand the behaviour of a mechanical two-dimensional mesh including rigid extended particles, let us start from a one-dimensional model, a chain composed of rigid rod-like particles and string connections, where $L'$ is the length of the rigid mass $M$, and $L$ the length of the string connecting two massive elements (see Fig.1). The vibrations of the chain we are considering are those perpendicular to the chain. In fact there are two possible directions, perpendicular to the chain: we consider just one direction because they are degenerate.

The unit cell of the lattice has a position given by the vector $(L+L')$. For simplicity, $L=L'$. The positions of the lattice sites (0) are denoted by the lattice indices $i, i+1, i+2, ...$, and the sites of the basis are denoted by $B$. The mass per unit length of the rigid rod is $\rho'$. Ropes have a linear density $\rho$. Due to the geometry, $T_o$ is the equilibrium axial force in each string line part.

Let us investigate the harmonic vibrations of the chain supposed to be infinite with displacements of lines and nodes along one of the transversal directions. $u_{i,b}$ is the displacement of one of the two nodes in the basis of the lattice cell at the reticular position $i$ from the equilibrium. $b$ can have then two possible determinations $0$ and $B$. With $w_{ji,B0}$, it is called the displacement of a string connecting a node $i$ in the lattice cell with the nearest neighbour node $j$. A linear co-ordinate $\zeta$ is ranging from zero to the length $L_{ji,B0}$ of the string. For the strings, the equation in the case of transverse vibrations is the usual wave equation, with a phase velocity $v = \sqrt{T_o/\rho}$. Solving the equation of motion for the string line units, we have as in Ref.15:

$$w_{ji,B0}(\zeta) = u_{i,B}\cos(\kappa\zeta) + \frac{u_{j,0} - u_{i,B}\cos(\kappa L)}{\sin(\kappa L)}\sin(\kappa\zeta) \tag{1}$$

and in the case of a time-harmonic oscillation of frequency $\omega$, $\kappa = \omega/v$.
The dynamic force the string exerts at node 0 equals:

$$f = -T_o \left.\frac{dw_{ji,B0}}{d\zeta}\right|_{\zeta=0} = T_o \frac{\omega}{v} \frac{u_{i,B}\cos(\kappa L) - u_{i,0}}{\sin(\kappa L)} \tag{2}$$

Assuming the rod-like particles in the chain as rigid, we can write the equations for the motion of the centre of mass and for rotation around it in a plane perpendicular to the chain. In the case of small displacements of the lattice nodes, equations are:

$$\frac{d^2(u_{i,B} + u_{i,0})}{dt^2} = \frac{T_o}{M}\left(\frac{dw_{ij,B0}}{d\zeta} + \frac{dw_{ij',B0}}{d\zeta}\right) \tag{3}$$

$$\frac{d^2(u_{i,B} - u_{i,0})}{dt^2} = \frac{L^2 T_o}{2I}\left(\frac{dw_{ij,B0}}{d\zeta} - \frac{dw_{ij',B0}}{d\zeta}\right) - \frac{LT_o}{2I}(u_{i,B} - u_{i,0}) \tag{4}$$

where index $j$ denotes the lattice site connected with the site $i$ on the right, and index $j'$ the site connected with $i$ on the left of the rigid particle. Let us remember that $dw_{ij,B0}/d\zeta$, $dw_{ij',B0}/d\zeta$ are proportional to the force components perpendicular to the chain, as from Eq.(2), and then pulling the rigid body. The force component parallel to the chain is simply considered equal to $T_o$. In Eq.(4), the first contribution on the r.h.s. represents the momentum of the dynamic forces $f$ on the rigid mass; the second contribution represents the momentum of axial tension $T_o$ on the rod-like particle. In Fig.1, a diagram shows the forces on the mass. Fig.1(b) illustrates that the axial forces produce a torque on the rod-like particle when it is not in the equilibrium position.

If we are looking for Bloch waves with wavevector k, it is possible to write for each lattice site:

$$u_{i,0} = u_0 \exp(i\omega t - 2ikL) \quad ; \quad u_{i,B} = u_B \exp(i\omega t - 2ikL) \tag{5}$$

and then the dispersion relations for the frequency $\omega$ can be easily obtained from the dynamic equations (3) and (4) of the rods. To simplify the solution, a Bogoliubov rotation of $u_0, u_B$ is possible. The Bolgoliubov transformation is the following:

$$u_0 = \eta + i\theta \quad ; \quad u_B = \eta - i\theta \tag{6}$$

and then we have explicitly the translation and the rotation degrees of freedom coupled in the two equations:

$$-M\omega^2 \eta = \frac{T_o}{vS}\{\eta\omega(\cos(kL)-C)+\theta\omega\sin(kL)\}$$

$$-2\frac{I}{L^2}\omega^2\theta = \frac{T_o}{vS}\{-\theta\omega(\cos(kL)+C)-S\theta+\eta\omega\sin(kL)\} \tag{7}$$

where $C = \cos(\kappa L)$; $S = \sin(\kappa L)$. For $\sin(\kappa L) = 0$, for any $\kappa$ standing wave, modes exist corresponding to internal vibration of the strings, with no associated nodal displacements. Let us consider the reduced frequency $\Omega = \omega/\omega_o$ where $\omega_o = T_o/Mv$. The dispersion relations of the chain as a function of the wavenumber $k$ are shown in Fig.2 for different values of the ratio $I/ML^2$. It is interesting to note the existence of a gap in the phonon dispersion between acoustic and optical modes.

**Honeycomb lattices**.
A two-dimensional model will be a planar membrane, for instance with a honeycomb structure, as shown in Fig.2 on the left: the lattice will be described using a unit cell with a convenient set of vectors ($\mathbf{l}_1, \mathbf{l}_2$), giving the lattice reticular positions. In the case of the honeycomb structure, the lattice has two nodes per unit cell. A mechanical model of the lattice can be made with rigid rod-like particles, with length $L$ and mass per unit length $\rho'$, and ropes with length $L$ and linear density $\rho$. The honeycomb lattice shown in Fig.2 on the right, has rigid connections between sites substituting all the bonds parallel to one of the lattice directions.
It is straightforward to investigate the harmonic vibrations of a two-dimensional mesh, if it is supposed to be infinite with displacements of lines and nodes in the direction perpendicular to its plane. As in the case of the one-dimensional lattice, $u_{i,b}$ is the displacement of one of the nodes in the basis of the lattice cell at the reticular position. The same for $w_{ji,bb'}$, the displacement of a string connecting a node in the lattice cell with the nearest neighbour node. If we are looking for Bloch waves with wavevector $\mathbf{k}$, it is possible to write for each lattice cell the displacements as:

$$u_{i,0} = u_0 \exp(i\omega t - i\mathbf{k}\cdot\mathbf{l}_1) \quad ; \quad u_{i,B} = u_B \exp(i\omega t - i\mathbf{k}\cdot\mathbf{l}_2) \tag{8}$$

If the basis has two sites for instance, the dispersion relations for the frequency $\omega$ are then obtained substituting these waves in the dynamics of rods, solving the following equations:

$$-\omega^2(u_{i,B}+u_{i,0}) = \frac{T_o}{M}\sum_{j,j'}\left(\frac{dw_{ij,B0}}{d\zeta}+\frac{dw_{ij',0B}}{d\zeta}\right)$$

$$-\omega^2(u_{i,B}-u_{i,0}) = \frac{L^2 T_o}{2I}\sum_{j,j'}\left(\frac{dw_{ij,B0}}{d\zeta}-\frac{dw_{ij',0B}}{d\zeta}\right)-\frac{L}{2I}\left(\sum_{j,j'}T_{o,\parallel}\right)(u_{i,B}-u_{i,0})$$

(9)

where $j, j'$ are indices for the nearest neighbour sites, as in the one-dimensional chain. The second term in the r.h.s. of the second equation is containing the components $T_{o,\parallel}$ of the forces, parallel to axial direction of the rod in its equilibrium configuration. For the honeycomb cell, $T_{o,\parallel} = T_o \cos(\pi/3)$.

In the case of the honeycomb lattice, there are two forces applied to the points 0 or $B$ of the rod: the contribution of the components of these two forces is positive and then giving a resulting torque stabilising the rod-like oscillators. Equation (9) becomes:

$$-\omega^2(u_{i,B}+u_{i,0}) = \frac{T_o}{M}\sum_{j,j'}\left(\frac{dw_{ij,B0}}{d\zeta}+\frac{dw_{ij',0B}}{d\zeta}\right)$$

$$-\omega^2(u_{i,B}-u_{i,0}) = \frac{L^2 T_o}{2I}\sum_{j,j'}\left(\frac{dw_{ij,B0}}{d\zeta}-\frac{dw_{ij',0B}}{d\zeta}\right)-\frac{LT_o}{2I}(u_{i,B}-u_{i,0})$$

(10)

The reduced frequency $\Omega = \omega/\omega_o$ (where $\omega_o = T_o/Mv$) of the honeycomb two-dimensional lattice can then be evaluated for different values of the ratio $I/ML^2$ as a function of the wavevector $\mathbf{k}$, in the directions OX, OC and OY, where O is the centre of the Brillouin Zone. The dispersions are shown in the Figure 4.

**Re-entrant honeycomb lattices**.
A two-dimensional model for an auxetic mechanical system is that proposed in Ref.[4] and shown on the left of Fig.5: the model was introduced to easily explain the behaviour of a material with a negative Poisson elastic parameter. The cell is a re-entrant honeycomb cell. The auxetic model can be viewed as a structure composed of rigid parts of length $L$ and $L'$: when the lattice is stretched, it expands instead to shrink. A model composed by ropes with length $L$ and rod-like particles $L'$ as in Fig.5 on the left is not suitable for discussing the vibrations, as we immediately explain.
Let us imagine to consider a two-dimensional honeycomb lattice where it is inserted in the honeycomb cell the rigid unit $L'$, as shown on the right part of Fig.5. On points 0 or $B$ of the rigid masses, three ropes and then three forces are acting. A force is parallel to the rod, then $T_{o,\parallel} = T_o$; the other two have components $T_{o,\parallel} = T_o \cos(2\pi/3)$, but negative with respect to the rod axis. If just these two forces were present in the model, as in a structure like Fig.5 on the left, the resulting torque in Eq.(9) would be destabilizing the acoustic oscillations of the membrane, in the case of a wavevector $\mathbf{k}$ in the X direction.
For the model in Fig.5 on the right, the dispersion relations for the frequency $\omega$ can be easily obtained from the dynamics of the rod-like particles. Fig.6 shows the reduced frequency as a function of the wavevector $\mathbf{k}$, in the directions OX, OC and OY, where O is the centre of the Brillouin Zone. In this image, the auxetic lattice has all the ropes subjected to the same tension $T_o$. The reduced frequency is $\Omega = \omega/\omega_o$, where $\omega_o = T_o/Mv$, as for the honeycomb lattice. The dispersion relations depend on the value of the ratio $I/ML'^2$. Note that $L'$ is the length of the rods, different from the length of the ropes. As previously told, the true auxetic network shown in Fig.5 on the left must be substituted with the auxetic lattice of Fig.5 on the right. We still call this model auxetic, because it is different from the honeycomb in the forces distribution.

Moreover, in this model, the strings parallel to the rigid masses can have a different tension $\xi T_o$. These ropes have a sound speed $v_2 = \sqrt{\xi T_o/\rho}$, different from the sound speed $v_1 = \sqrt{T_o/\rho}$ of the other ropes. The reduced frequency we use is $\Omega = \omega/\omega_o$, with $\omega_o = T_o/Mv_1$. The Eq.(9) must be rewritten in the following form:

$$-\omega^2 \left(u_{i,B} + u_{i,0}\right) = \frac{T_o}{M} \sum_{j,j'} \left(\frac{dw_{ij,B0}}{d\zeta} + \frac{dw_{ij',0B}}{d\zeta}\right) + \frac{\xi T_o}{M} \sum_{k,k'} \left(\frac{dw_{ik,B0}}{d\zeta} + \frac{dw_{ik',0B}}{d\zeta}\right)$$

$$-\omega^2 \left(u_{i,B} - u_{i,0}\right) = \frac{L'^2 T_o}{2I} \sum_{j,j'} \left(\frac{dw_{ij,B0}}{d\zeta} - \frac{dw_{ij',0B}}{d\zeta}\right) + \frac{L'T_o}{2I} \left(u_{i,B} - u_{i,0}\right) + \tag{11}$$

$$+ \frac{L'^2 \xi T_o}{2I} \sum_{k,k'} \left(\frac{dw_{ik,B0}}{d\zeta} - \frac{dw_{ik',0B}}{d\zeta}\right) - \frac{L'\xi T_o}{2I} \left(u_{i,B} - u_{i,0}\right)$$

where indices $j, j'$ are used when sites are connected by ropes with tension $T_o$ and $k, k'$ when sites are connected with ropes with tension $\xi T_o$. Note the different sign from Eq.10. An approach to solve system (11) with the Bogoliubov transformation was previously proposed [16].

The Fig.6 shows the phonon dispersions of the auxetic lattice, for different values of ratio $I/ML'^2$ and for $\xi = 1$. When parameter $\xi$ increases over a certain value, a complete bandgap between the acoustic and the optical mode appears and this is shown in Fig.7. Of course, this approach with different axial tensions is possible for the honeycomb model too.

In both models, honeycomb and auxetic, the rod-like masses are viewed by the waves in the Y-direction as point-like ones.

**Auxetic membranes.**
We have seen that a complete bandgap is easily obtained in two-dimensions, adjusting lattice parameters and interactions, that is changing elastic properties or densities of ropes. Of course, different and more complex auxetics must be proposed and studied, to exhaustively understand the behaviour of these structures. If we consider as "auxetic-like" two-dimensional structures, those structures which do not collapse, when stretched along one of the in-plane directions, several membranes can be proposed, but it is necessary to insert some rigid parts in their mesh.

Let us consider for instance the square lattice in the upper part of Fig.8. The thick lines represents the rod-like particles, which have different orientations in the plane of the lattice. Then the lattice unit contains two rigid rods. In the lower part of the figure, the phonon dispersions are shown. If the masses in the lattice unit are different, a complete band-gap appears, in agreement with the behaviour of crystalline systems [3] and mechanical systems with point-like masses proposed in Ref.[15].

More complex two-dimensional structures can be proposed as that shown in Fig.9. The structure resembles the model for auxetics proposed in [17] and the membrane studied in [18]. This last reference discusses the in-plane vibrations of rectangular rigid particles connected by harmonic elements.

Of course, different approaches to the problem of vibrations of auxetic structures are possible. For instance, a solution based on finite elements was used in Ref. [19] to solve a macroscopic mechanical system. These studies are in fact very important for applications. The aim of this paper is instead the investigation of the role played by rod-like particles in the lattice vibrations and if it is possible to create a band-gap with proper mass differences or interaction anisotropy in the unit cell of the lattice.

**Figure captions**

Fig.1:  The chain with rigid particles in the upper part ( $L$ is the length of the rods and  strings). In (a), the dynamic forces acting on the rigid particle and in (b) the axial tension of ropes.

Fig.2:  Reduced frequency of the phonon dispersions as a function of the wavelength for different values of the ratio $I/ML^2$ (0.1 red, 1 green and 10 blue). Note the behaviour of the optical mode for high values of the ratio $I/ML^2$. The gap between acoustic and optical modes is quite pronounced for the green curve.

Fig.3: The honeycomb structure. On the right, the primitive lattice points (black dots) and the points of the basis (white dots), and a set of two lattice vectors  convenient for calculations. Thick lines are representing the rigid rods.  The three directions OX, OC and OY along which  evaluate the dispersion relations  are also displayed.

Fig.4: The phonon dispersions for the honeycomb lattice  for three different ratios $I/ML^2$ (a=0.5, b=1, c=5). The group velocity is strongly dependent on the direction. For the Y direction, the acoustic mode does not change when the ratio $I/ML^2$ changes. In fact, the rod-like masses are viewed by the waves in this direction as point-like masses.

Fig.5: The auxetic structure. For the auxetic lattice the Poisson coefficient is negative: if the lattice is stretched, it expands instead to shrink. $L$ and $L'$ are the lengths of the rigid rods in the auxetic mesh. On the right, the mesh used to investigate vibrations. The figure shows the primitive lattice points (black dots) and the points of the basis (white dots), and a set of two lattice vectors convenient for calculations. Thick lines are representing the rigid rods, thin lines the ropes. Note that at each point 0 or B are attached three ropes. The directions OX, OC and OY along which the dispersion relations are evaluated are also displayed.

Fig.6: The phonon dispersions for the auxetic lattice for three different ratios $I/ML'^2$ (a=0.5, b=1, c=5). Parameter $\xi = 1$. For the Y direction, the acoustic mode does not change when the ratio $I/ML'^2$ changes.

Fig.7: The phonon dispersions for the auxetic lattice with ratio $I/ML'^2 = 1$ for $\xi = 4$ (a), and $\xi = 1$ (b) reported for comparison. Note the complete band-gap for (a). The same result is possible in the honeycomb lattice.

Fig.8: The phonon dispersions for the square lattice depicted in the upper part. If the rods parallel to X direction have a different mass from the rods parallel to Y, then a complete band-gap appears (mass ratio 2, red lines).

Fig.9: Phonon reduced frequency for the lattice depicted in the upper part. If the rods parallel to X direction have a different mass from the rods parallel to Y, then a complete band-gap appears (mass ratio 4).

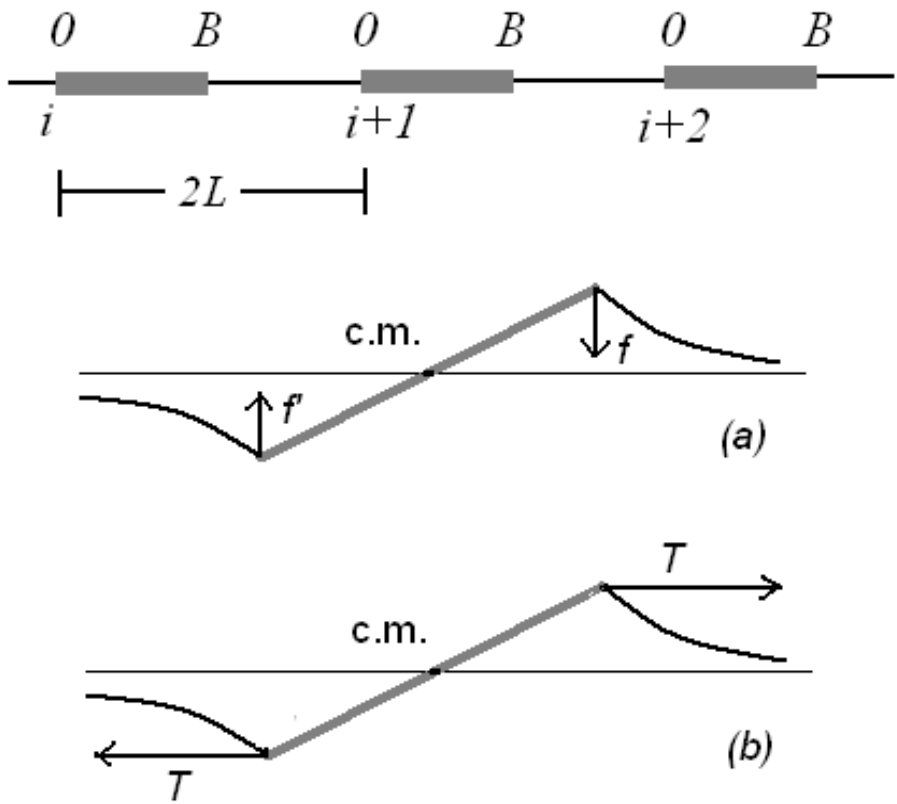

Fig.1

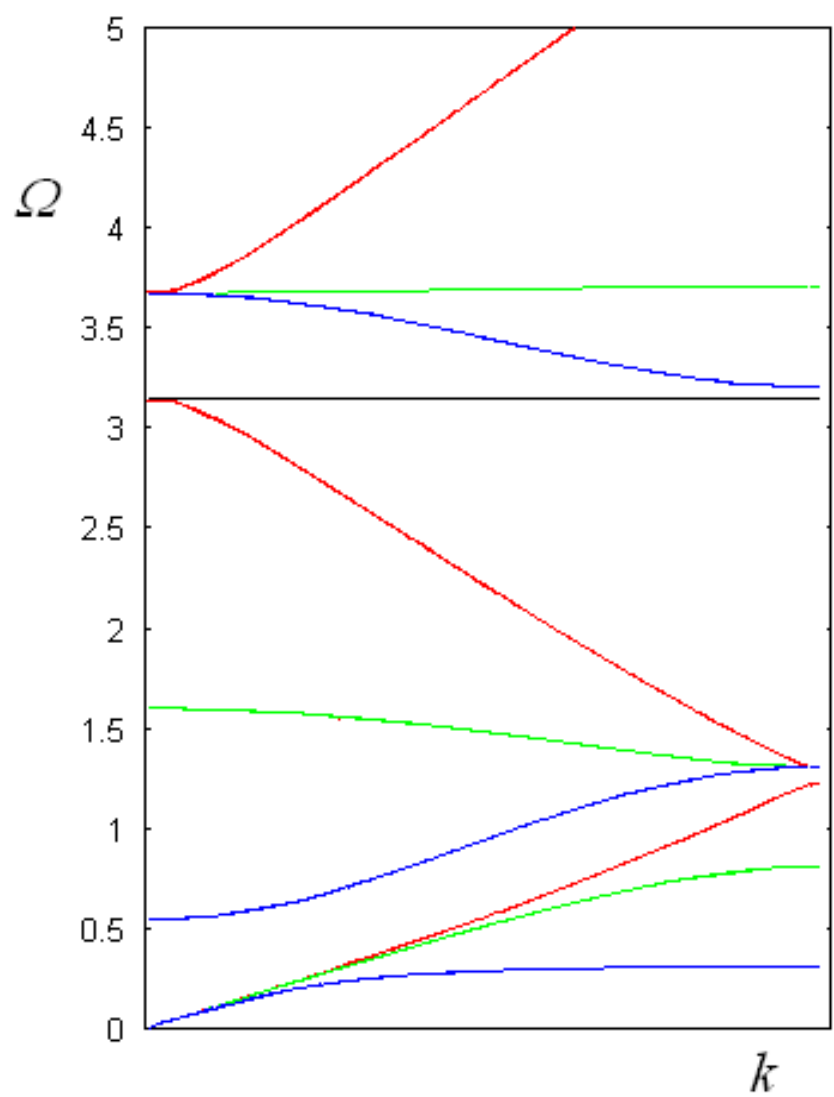

Fig.2

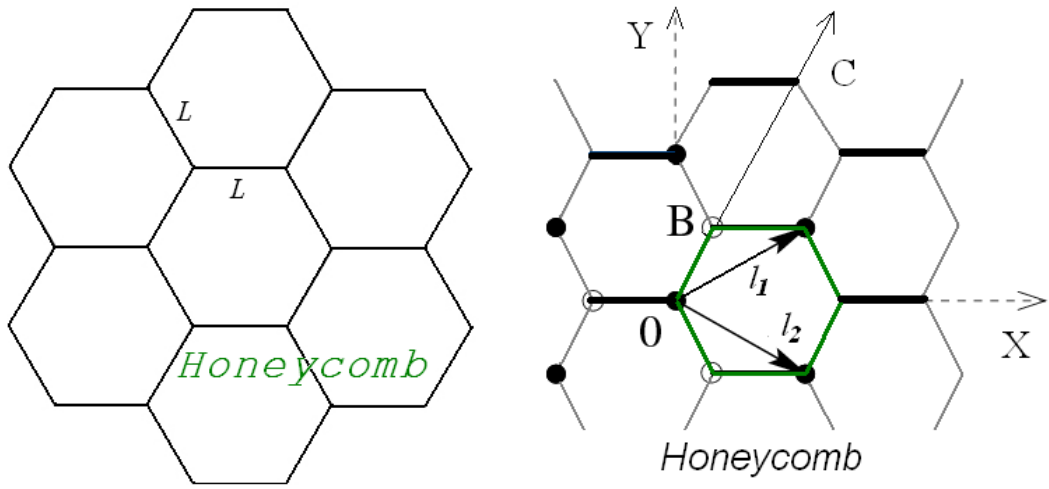

Fig.3

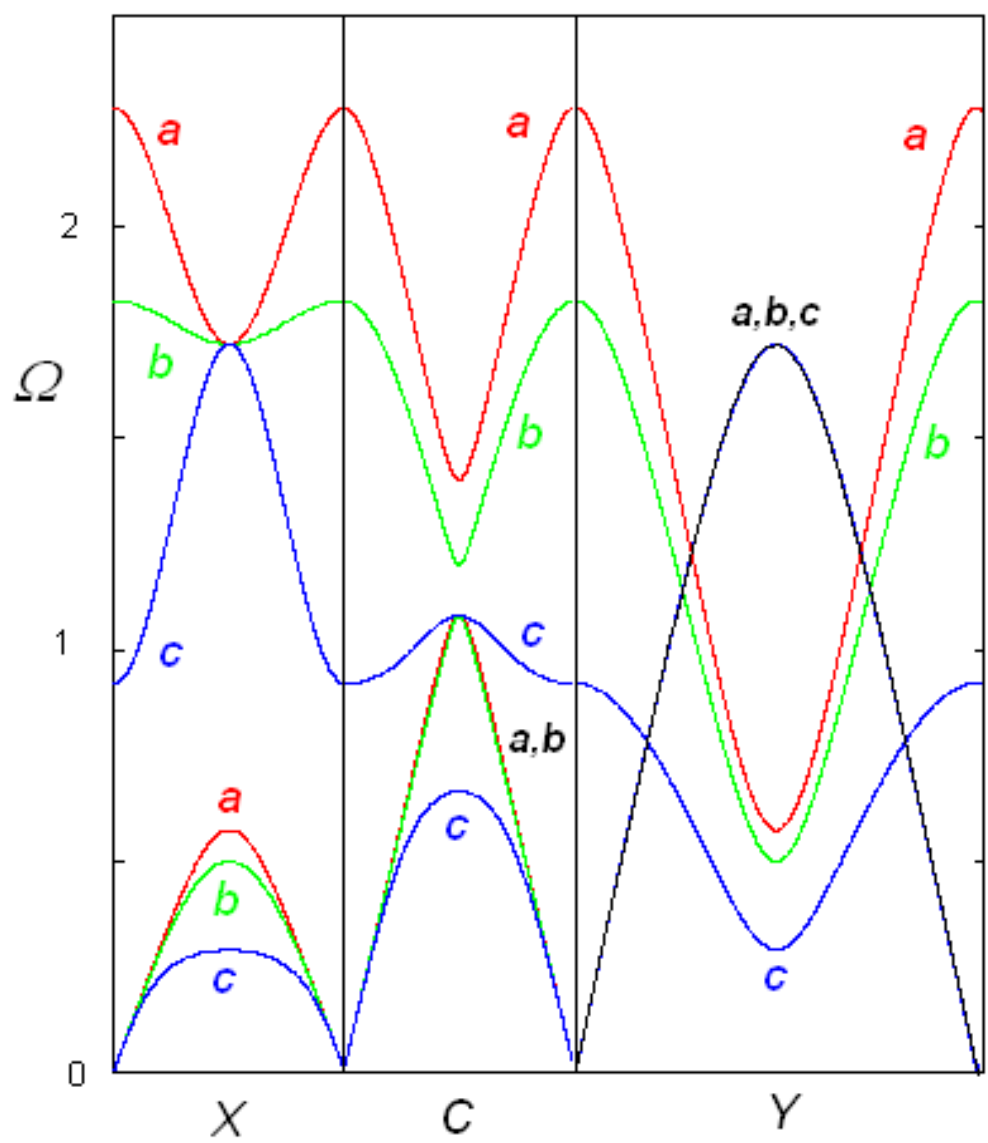

Fig.4

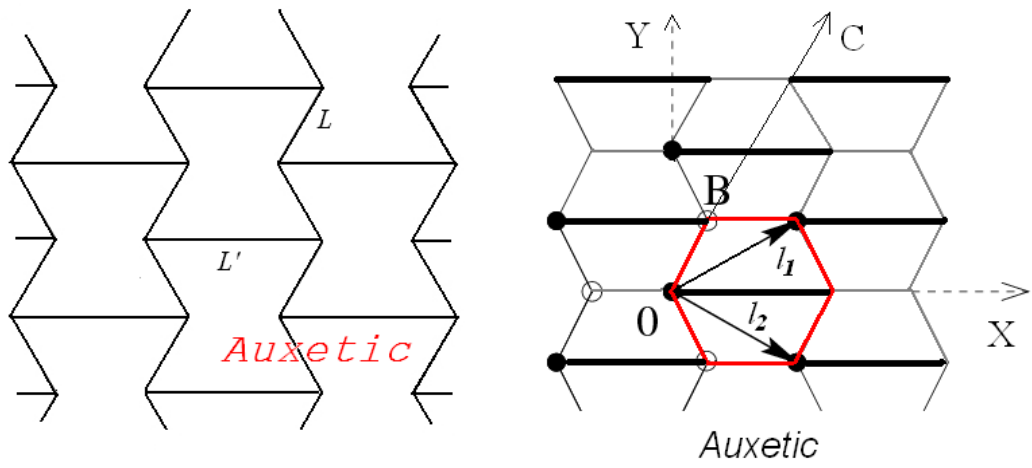

Fig.5

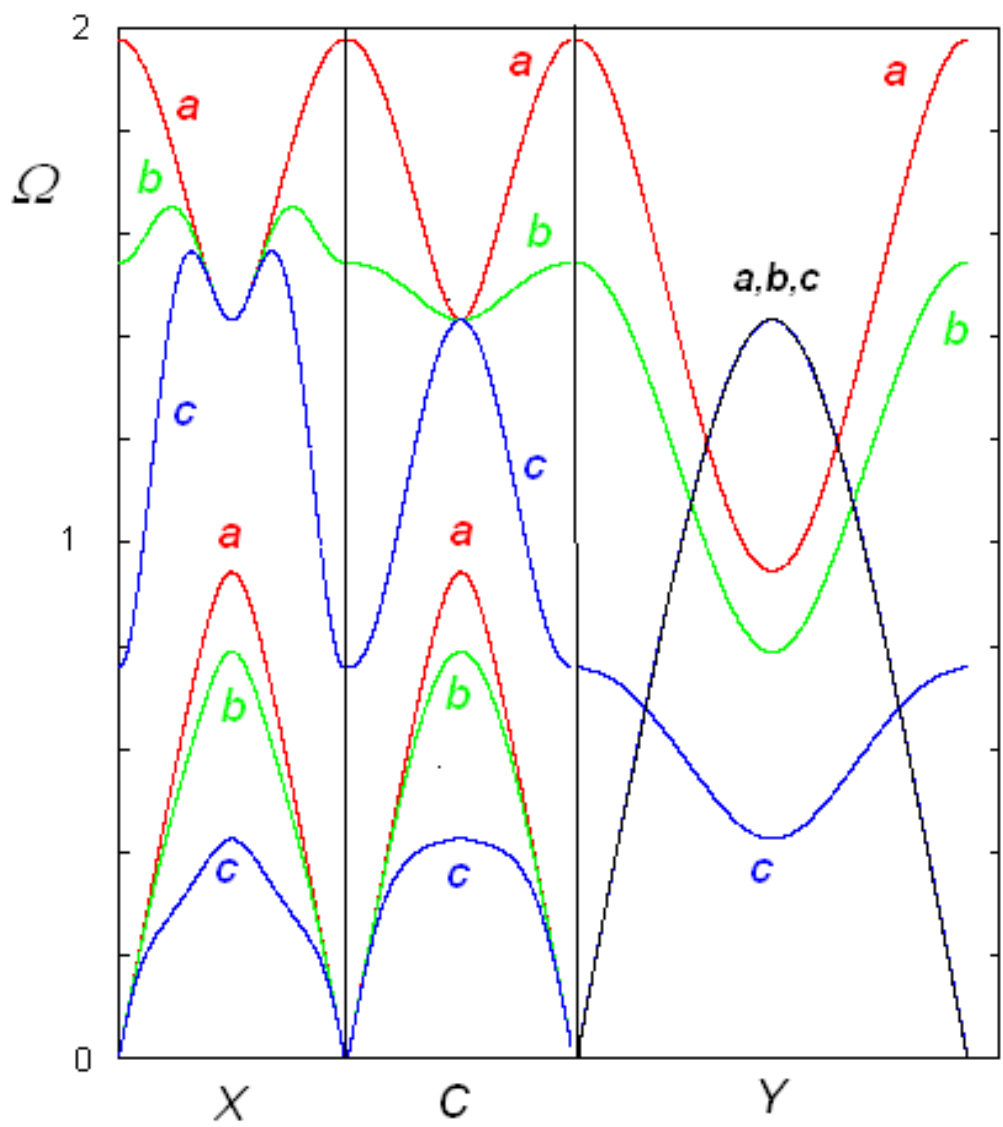

Fig.6

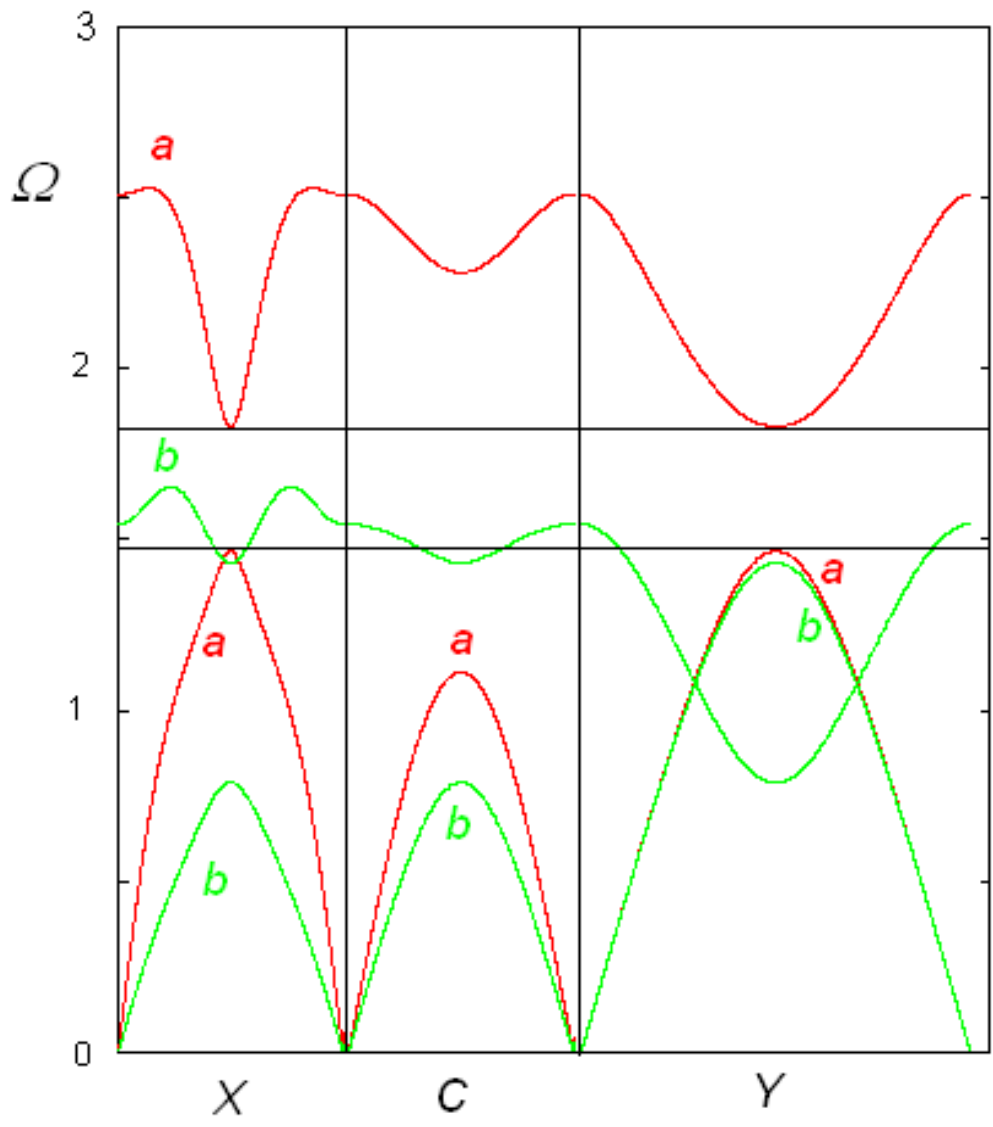

Fig.7

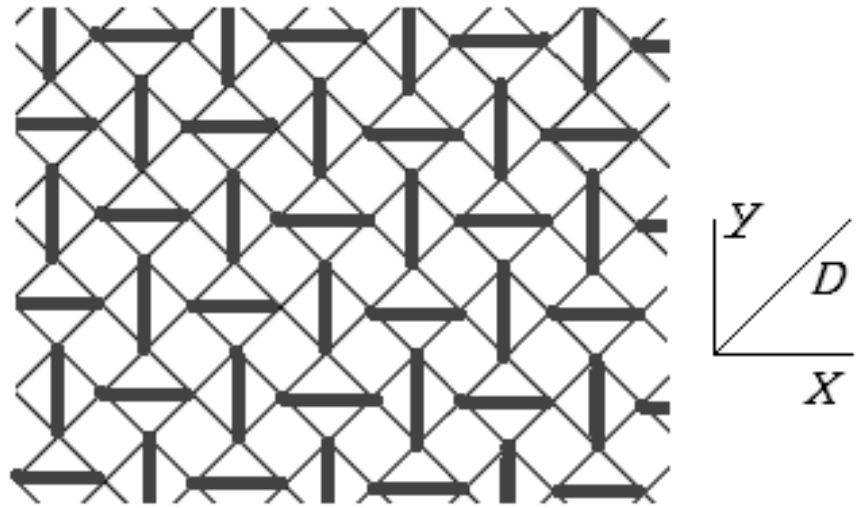

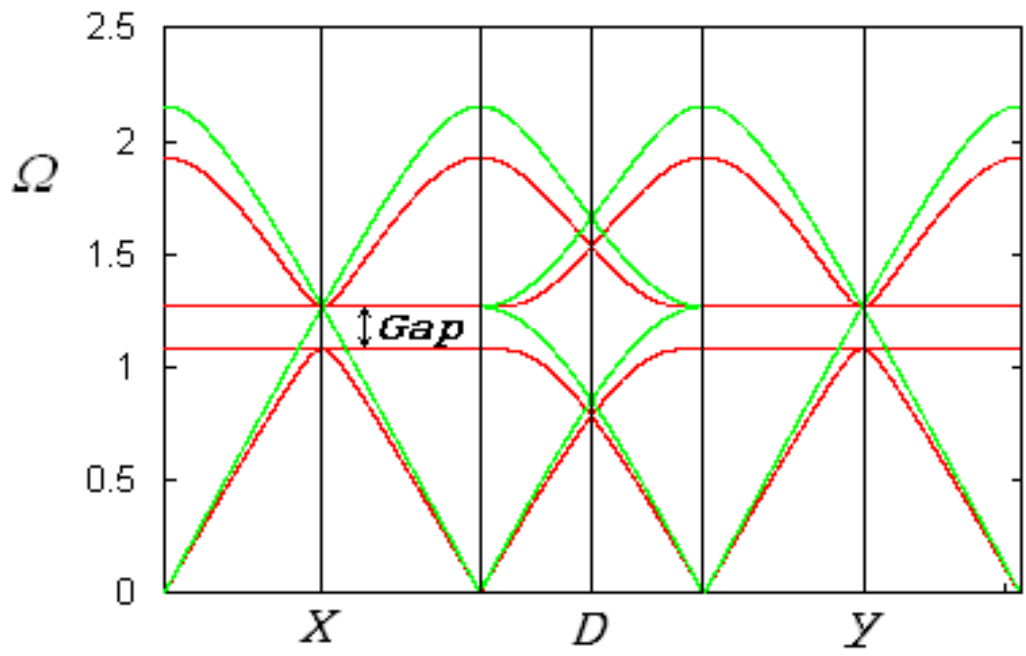

Fig.8

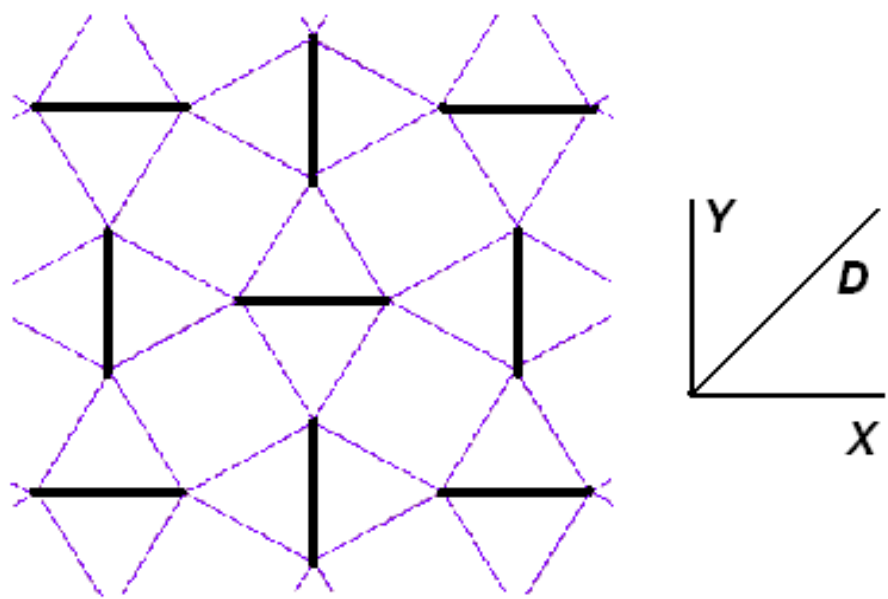
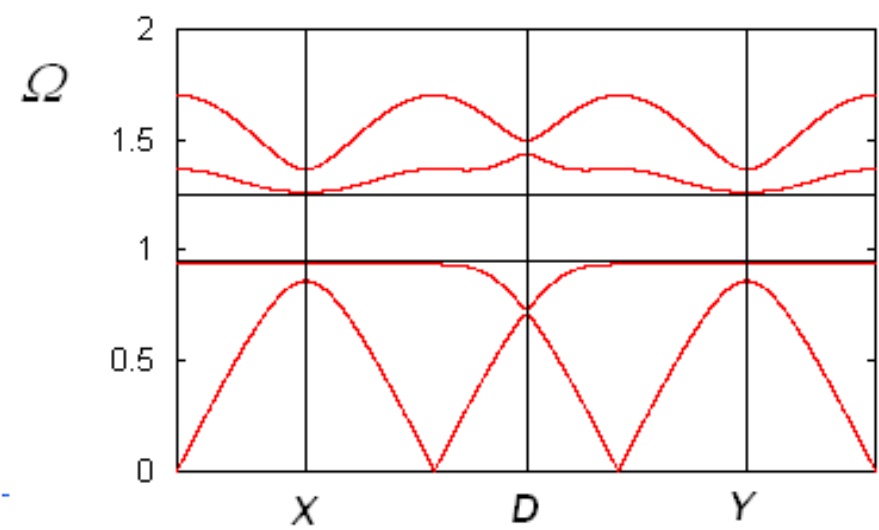

Fig.9